\documentclass{aa}
\usepackage{epsfig}
\begin{document}
\thesaurus{06(06.05.01; 06.09.1; 06.15.1)}
\renewcommand{\dbltopfraction}{1}
\renewcommand{\textfraction}{0}
\def\figdir{./fig}
\def\be{\begin{equation}}
\def\ee{\end{equation}}
\def\bearr{\begin{eqnarray}}
\def\eearr{\end{eqnarray}}
\def\barr{\begin{array}}
\def\earr{\end{array}}
\def\p{\partial}
\def\dis{\displaystyle}
\def\apj{ApJ }
\def\mnras{MNRAS }
\def\dd{{\rm d}}
\def\alphac{\alpha_{\rm c}}
\def\nablaad{\nabla_{\rm ad}}
\def\nablarad{\nabla_{\rm rad}}
\def\Xs{X_{\rm s}}
\def\pcz{p_{\rm cz}}
\def\rcz{r_{\rm cz}}
\def\dcz{d_{\rm cz}}
\def\kappat{\tilde \kappa}
\def\deltar{\delta_r}
\def\cm{\,{\rm cm}}
\def\erg{\,{\rm erg}}
\def\dyn{\,{\rm dyn}}
\def\s{\,{\rm s}}
\def\note #1]{{\bf #1]}}
\def\etal{{\it et al.}}
\def\gwig{{\leavevmode\kern0.3em\raise.3ex\hbox{$>$}
\kern-0.8em\lower.7ex \hbox{$\sim$}\kern0.3em}}
\def\lwig{{\leavevmode\kern0.3em\raise.3ex\hbox{$<$}
\kern-0.8em\lower.7ex \hbox{$\sim$}\kern0.3em}}

\title{Opacity effects on the solar interior}
\subtitle{I. Solar structure}

\author{S. C. Tripathy\inst{1}
\and
J. Christensen--Dalsgaard\inst{2,3}}
\offprints{S. C. Tripathy}
\institute{Udaipur Solar Observatory, Physical Research Laboratory.
PO Box No. 198, Udaipur 313 001, India\\
e-mail: sushant@uso.ernet.in
\and
Teoretisk Astrofysik Center, Danmarks Grundforskningsfond
\and
Institut for Fysik og Astronomi, Aarhus Universitet,
DK--8000 Aarhus C, Denmark\\
e-mail: jcd@obs.aau.dk}
\date{Received date; accepted date}
\maketitle
\begin{abstract}
Despite recent major advances, the opacity remains a source
of substantial uncertainty in the calculation of solar models,
and hence of solar oscillation frequencies.
Hence it is of substantial interest to investigate the
sensitivity of solar structure to changes in the opacity.
Furthermore, we may hope from the precise helioseismic inferences
of solar structure to obtain information about possible 
corrections to the opacities used in the model calculation.
Here we carry out detailed calculations of the influence on
solar models of changes in the opacity, including also
evolutionary effects.
We find that over the relevant range the response of the model
is approximately linear in the opacity change, allowing the
introduction of \emph{opacity kernels} relating a general
opacity change to the corresponding model changes.
Changes in the convection zone can be characterized
entirely by the change in the initial composition and mixing length
required to calibrate the model.
\end{abstract}
\section{Introduction}
Accurate frequency measurements of thousands of
modes of solar acoustic oscillation 
provide detailed information on the solar interior. 
In order to use
these frequencies to derive internal structure and dynamics of
the Sun, it is crucial to understand and limit the uncertainties
in the computation of solar models and mode frequencies. 
Furthermore, it is of considerable interest to investigate
the sensitivity of solar structure 
to changes in the input physical parameters and properties of
the solar interior.
One of the important physical properties in
solar model calculations is the opacity,
which is intimately
linked to the properties of the solar oscillations through its
effects on the mean structure of the Sun. 
Here we investigate the sensitivity of solar structure to
local modifications to the opacity.

Several authors have investigated the effect of opacity on the 
solar models and oscillation frequencies.
In an early paper, 
Bahcall, Bahcall \& Ulrich (1969) studied the sensitivity of the solar neutrino
fluxes to localised changes in the opacity and equation of state and 
concluded that the neutrino capture rates are more sensitive to 
the equation of state.
The effects of artificial opacity modifications
on the structure of solar models 
and their frequencies 
were also examined by Christensen-Dalsgaard (1988).
Constructing static models of the present Sun with an enhanced opacity 
near the base of the convection zone, he showed that the changes 
in structure and frequencies are approximately 
linear even to an opacity change of 60\%. 
The linearity of the response 
of the model to opacity changes was later confirmed by 
Christensen-Dalsgaard \& Thompson (1991).
In a detailed investigation, Korzennik \& Ulrich (1989) attempted 
to improve the agreement
between the theoretical and observed frequencies of oscillations by 
determining corrections to the opacity through inverse analysis.
They found that the opacity inversion can only partially resolve the 
discrepancy.
In a similar analysis, Saio (1992) obtained opacity 
corrections by fitting to low-degree frequency separations
and helioseismically inferred sound speeds at a few points in the model;
he found that much of the discrepancy between the Sun and the model
could be removed by opacity changes of up to 50 \%.
More recently, Elliott (1995) investigated the helioseismic
inverse problem as expressed in terms of corrections to the opacity.
He derived kernels relating the opacity differences to the changes
in frequencies, on the assumption that the change in luminosity could
be ignored; he proceeded to carry out inversions for the opacity
errors, based on the observed solar oscillation frequencies,
and neglecting possible changes in composition associated with
the change in opacity.
He found that the differences between observed and computed oscillation
frequencies could be accounted for by opacity changes of up to about 5 \%.

Here we carry out a detailed investigation of the
sensitivity of solar structure to localised changes in the opacity,
both for static and evolutionary solar models.
The ultimate goal, taken up in a subsequent paper, is
to determine the opacity corrections required to account for
the helioseismically inferred properties of the solar interior.
In general the opacity $\kappa$ is a function of
density $\rho$, temperature $T$ and composition $\{X_i\}$.
However, it is evident that information about the properties
of the present Sun cannot constrain the opacity in such generality. 
Thus, for simplicity, we consider only opacity modifications
that are functions $\delta \log \kappa(T)$ of temperature alone,
$\log$ being the logarithm to base 10.
If the opacity correction is sufficiently small that higher-order
terms can be neglected, 
the response of any quantity $F$ related to solar structure can be
expressed in terms of a differential kernel $K_F$ as 
\be
{\delta F \over F} = \int K_F(T)~\delta\log\kappa(T)~\dd \log T \; .
\label{kernel}
\ee

Here for simplicity we estimate the kernels from
\be
K_F(T_0) = {(\delta F/F)(T_0)
\over \int \delta\log\kappa(T) \dd \log T} \; ,
\label{kdef}
\ee
where $\delta F$ is the change corresponding to a 
suitably change $\delta \log \kappa$ localised at $T = T_0$.
Relations similar to Eq.~(\ref{kernel}) form the 
basis of inverse analysis (Gough 1985):
if $\delta F$ is the difference between the observed
and theoretical frequencies, 
Eq.~(\ref{kernel}) can in principle be inverted to determine corrections to the
opacity in the model
(Korzennik \& Ulrich 1989; Saio 1992; Elliott 1995).
The kernels also provide a powerful visualization of the
response in a given physical quantity of the solar model 
to a small perturbation in the input physics. 
As an example of this we evaluate kernels  
relating opacity changes to
the structural differences in the solar model,
the neutrino fluxes,
and the small frequency separation between low-degree modes.

\section{Procedure}
To illustrate the sensitivity of the models to opacity changes
we have computed extensive sets of comparatively simple models.
These were based on the OPAL opacity tables (Iglesias, Rogers \& Wilson 1992),
the Eggleton, Faulkner \& Flannery (1973) equation of state,
and nuclear reaction parameters from Parker (1986) 
and Bahcall \& Ulrich (1988).
Convection was treated with the mixing-length formulation
of B\"ohm-Vitense (1958).
The neutrino capture rates for the $^{37}$Cl and $^{71}$Ga experiments 
were obtained with the cross-sections given by Bahcall (1989).
The heavy-element abundance $Z$ was 0.01895.
All models were calibrated to a fixed radius ($6.9599 \, 10^{10} \cm)$
and luminosity ($3.846 \, 10^{33} \erg \s^{-1}$)
at the assumed solar age of $4.6 \, 10^9$ years
by adjusting the mixing-length parameter $\alphac$ and the composition.
Diffusion and gravitational settling were ignored.
Details of the computational procedure were
described by Christensen-Dalsgaard (1982).

It should be noted that our models are somewhat simplified,
particularly in the choice of equation of state and the neglect
of settling, compared with present state-of-the-art solar models
(for a review see, for example, Christensen-Dalsgaard {\etal} 1996).
Thus the values of the model parameters quoted, e.g.,
in Table 1 cannot be regarded as representative of the actual solar structure.
However, we are here concerned with the {\it differential}
effects on the models of changes to the opacity;
for this purpose, the present simplified physics is entirely adequate.

We have considered two types of models: 
static models of the present Sun and proper evolution models, evolved from a
chemically homogeneous initial Zero-Age Main-Sequence (ZAMS) model.
In the static models, the hydrogen profile $X(q)$ 
where $q$ = $m/M$ ($m$ being the
mass interior to the given point and $M$ is the total mass of the Sun), 
was obtained by scaling by a constant factor the profile $X_{\rm r}(q)$,
obtained from a complete evolution model:
\be
X(q) = \chi X_{\rm r}(q) \; ,
\label{Xscale}
\ee
where $\chi$  was adjusted to fit the solar luminosity.
In the evolution models the calibration to solar luminosity
was achieved by adjusting the initial helium abundance $Y_0$.

The sensitivity of solar structure to opacity was investigated by 
increasing $\kappa$ 
in a narrowly confined region near a specific temperature 
$T_0$ according to
\be
\log \kappa = \log \kappa_0 + f(T_0) \; ,
\ee
where $\kappa_0$ is the opacity as obtained from
the opacity tables;
$T_0$ was varied over the temperature range of the model.
The function $f(T_0)$ has the form
\be
f(T_0)
= A \exp \left[ - {\left(\log~T~-~\log~T_0\over \Delta\right)}^2 \right] \; ,
\label{kapmod}
\ee
where the constants $A$ and $\Delta$ determine the magnitude and width of the 
opacity modification.
Henceforth the models computed with the same input physics but without
opacity modifications will be referred to as reference models whereas the
perturbed ones will be referred as modified models. 

\begin{table*}
\caption[]{Properties of static solar models. Here
$X_0$ is the initial hydrogen abundance, 
$\alphac$ is the mixing-length parameter,
$d_{\rm cz}/R$ is the depth of the
convection zone in units of photospheric radius $R$,
$T_{\rm c}$ and $X_{\rm c}$ are central temperature and
hydrogen abundance of the model of the present Sun,
and the last three columns give the neutrino capture rates
in the ${}^{37}$Cl and ${}^{71}$Ga experiments, and the 
flux of ${}^8$B neutrinos.
}
\begin{flushleft}
\begin{tabular}{clcccccccc}
\hline
&&&&&&&\multicolumn{3}{c}{Neutrino flux}\\
Model&$A$&$X_0$&$\alphac$&$d_{\rm cz}/R$&$T_{\rm c}$&$X_{\rm c}$&
$^{37}$Cl&$^{71}$Ga&$^8$B\\
&&&&&$(10^6 {\rm K})$&&(SNU)&(SNU)&$10^6 \cm^{-2} \s^{-1}$\\
\hline
S0&0.0&0.6978&1.8098&0.2739&15.604&0.3383&8.092&132.02&5.747\\
SA&0.02&0.6972&1.8038&0.2733&15.601&0.3380&8.073&131.97&5.730\\
SB&0.1&0.6948&1.7813&0.2709&15.589&0.3369&8.006&131.84&5.670\\
SC&0.2&0.6920&1.7561&0.2681&15.576&0.3355&7.939&131.72&5.608\\
\hline
\end{tabular}
\end{flushleft}
\end{table*}

The detailed response of the oscillation frequencies to the opacity
modifications will be considered separately.
However, the importance of the small frequency separation as 
a diagnostics of the solar core motivates that we include it here, 
for comparison with the response of the neutrino flux.
We characterize the separation,
averaged over radial order $n$ and degree $l$,
by the parameter $D_0$ defined by
\be
\langle \nu_{n{\ell}} - \nu_{n-1,{\ell}+2}\rangle_{n,l}
\simeq (4{\ell} + 6)D_0 \; ,
\label{freqsep}
\ee
where asymptotically
\be
D_0 \simeq 
- {1\over{4\pi^2(n_0+1/4+\beta)}}
\int_0^R {\dd c\over \dd r}{\dd r\over r} \; .
\label{d0}
\ee
Here $n_0$ is a suitable reference order, $\beta$ is a constant
predominantly related to the structure of the outer parts of the model,
and $c$ is the sound speed.
Details of the calculation of the oscillation
frequencies and the evaluation of the average in
Eq. (\ref{freqsep}) were given by
Christensen-Dalsgaard \& Berthomieu (1991).
The fit to the frequencies included modes with $l = 0 - 4$
and radial order such that $17 \le n + l/2 \le 29$;
$n_0$ was chosen to be 23.
%

\section{Results}

\subsection{Linearity of the response}
The linearity of the response of model parameters 
to the opacity modifications was tested 
by constructing  static models with $A = 0.02, 0.1$ and 0.2
(note that $A = 0.1$ corresponds to a maximum change in opacity of 26\%).
The modifications were centred at $\log T_0 = 7$, 
and the width was $\Delta = 0.02$
(here and in the following $T$ and $T_0$ are measured in K).
The properties of the reference and modified models 
are presented in Table~1. 
It is evident 
that the changes in opacity required fairly substantial changes in the 
composition and mixing-length parameter to calibrate the model to obtain 
the correct luminosity and radius, and led to a significant change in the 
depth of the convection zone.  

Fig.~1 shows the change in sound speed, temperature and density 
resulting from the opacity 
change with $A$ = 0.1 and 0.2, the differences for $A$ = 0.2 being scaled
by 1/2 to test linearity.
Even though the curves do not coincide 
precisely, the model changes are approximately linearly related
to even quite substantial changes in $\log \kappa$, as 
was also found in earlier investigations.
We have in addition confirmed that the changes are relatively insensitive
to the width $\Delta$, provided the integrated modification
$\int \delta \log \kappa \, \dd \log T$ is kept fixed.

\begin{figure}
\vspace{0cm}
\hspace{0cm}\epsfxsize=8.8cm \epsfbox{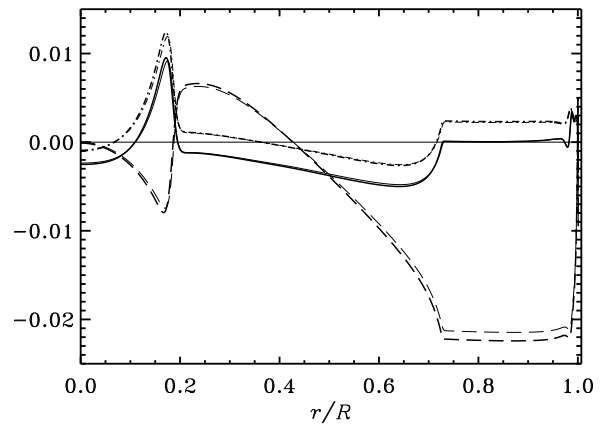}
\vspace{0cm}
\caption[]{Differences at fixed fractional radius $r/R$ between
static modified models SB and SC and the reference model S0, 
in the sense (modified model) -- (reference model).
The heavy lines show results for $A = 0.1$ and the thin lines 
results for $A = 0.2$, multiplied by $1/ 2$ to
illustrate the extent to which the response is linear.
The following differences are illustrated:
$\deltar \ln c^2$ (solid line);
$\deltar \ln \rho$ (long-dashed line); and
$\deltar \ln T$ (dash-dotted line),
$\ln$ being the natural logarithm
}
\end{figure}

A requirement for linearity is that the model does not
become convectively unstable at the location where the
opacity is modified.
This constraint can approximately be formulated
as
\be
(\delta \ln \kappa)_{\rm max} = (\ln 10) \, A < {\nablaad \over \nabla} - 1 \; ,
\label{lim}
\ee
where $\nabla = \dd \ln T/\dd \ln p$ and $\nablaad$ is its adiabatic
value, both evaluated in the reference model.
We note that this constraint is only barely satisfied for $A = 0.2$
and $\log T_0 = 7$: had we used $A = 0.3$ instead, the region
around the modification would have become convectively unstable
and the structure of the model would have been drastically altered.

In the following we consider modifications computed
using $A = \Delta = 0.02$, corresponding to Model SA in Table~1;
this is sufficiently localized to represent the linear response of
opacity modifications confined to very narrow local regions, 
similar to delta functions.
Furthermore, this value of $A$ ensures that the constraint (\ref{lim})
is satisfied, except in a few cases in the core 
during early stages of evolution,
of little significance to the structure of the model of the present Sun.
Specifically, a small short-lived initial convective core
appeared in the evolution calculations for $7.11 < \log T_0 < 7.15$; 
to avoid problems with non-linearities we simply suppressed the
mixing of the composition of the core in these cases.

\subsection{Model differences} 
\label{moddif}
\begin{figure}
\vspace{0cm}
\hspace{0cm}\epsfxsize=8.8cm \epsfbox{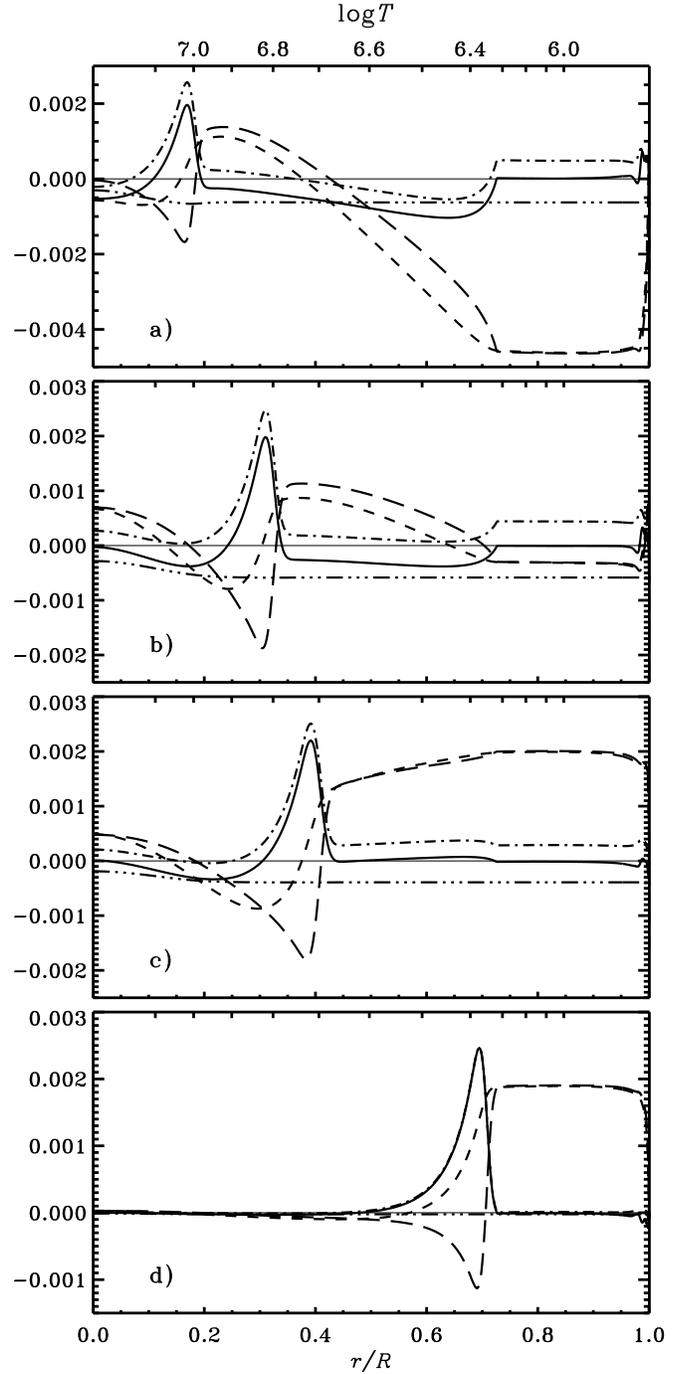}
\vspace{0cm}
\caption[]
{Differences between modified static models
and the reference  model S0 in the sense (modified model) -- (reference model).
The modified models correspond to
the following locations of the opacity increases:
{\bf a} $\log T_0 = 7.0$;
{\bf b} $\log T_0 = 6.8$;
{\bf c} $\log T_0 = 6.7$;
{\bf d} $\log T_0 = 6.36$.
The variables shown are:
$\deltar \ln c^2$ (solid line);
$\deltar \ln p$ (short-dashed line);
$\deltar \ln \rho$ (long-dashed line);
$\deltar \ln T$ (dash-dotted line); and $\deltar X$ (dash-triple-dotted line)
}\label{statdiff}

\end{figure}

\begin{figure}
\vspace{0cm}
\hspace{0cm}\epsfxsize=8.8cm \epsfbox{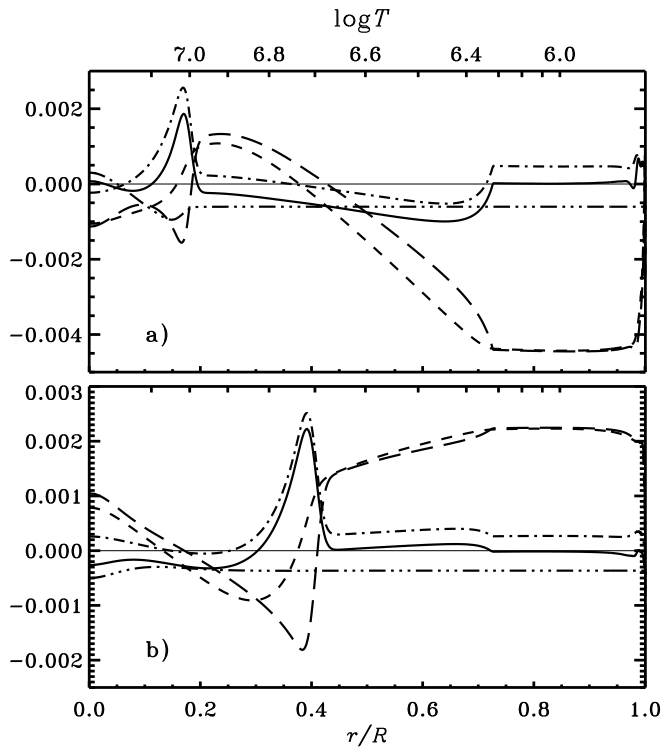}
\vspace{-8cm}
\caption[]{
Differences between modified evolution models
and the corresponding reference model,
in the sense (modified model) -- (reference model).
The modified models correspond to
the following locations of the opacity increases:
{\bf a} $\log T_0 = 7.0$;
{\bf b} $\log T_0 = 6.7$.
The line styles have the same meaning as in Fig.~\ref{statdiff}
}\label{evoldiff}
\end{figure}

\begin{figure}
\vspace{0cm}
\hspace{0cm}\epsfxsize=8.8cm \epsfbox{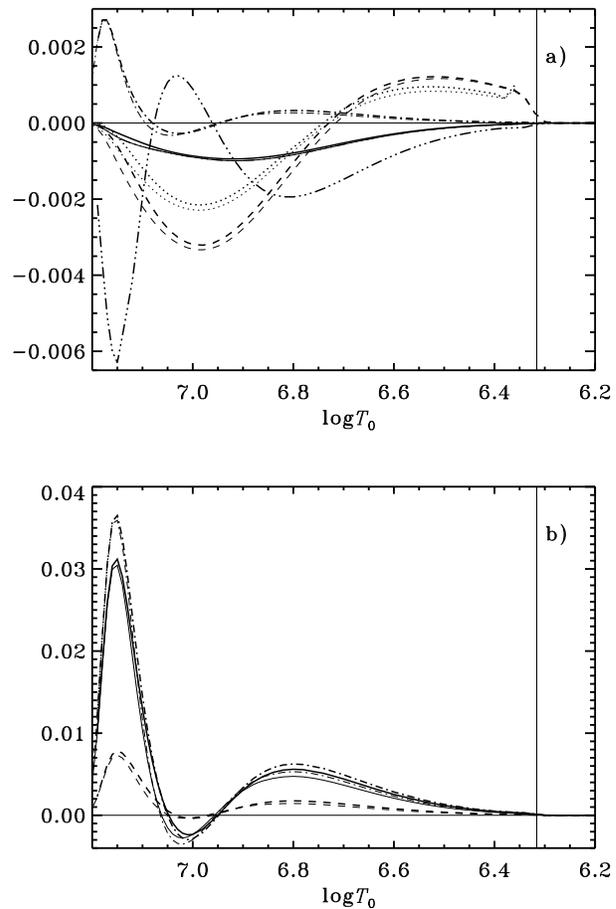}
\vspace{0cm}
\caption[]{
Responses of various global model quantities, as a function of the location
$T_0$ of the opacity modification,
shown in the sense (modified model) -- (reference model).
Thin lines correspond to static models and heavy lines to evolution models.
The thin vertical line corresponds to the temperature at
the base of the convection zone in the reference model.
{\bf a} Relative changes
in central temperature $T_{\rm c}$ (dash-dotted line), 
depth of the convection zone (dotted line), and
mixing-length parameter $\alphac$ (dashed line),
initial hydrogen abundance $X_0$ (solid line)
and central hydrogen abundance $X_{\rm c}$ (dash-triple-dotted line).
(Note that for the static models, the relative changes in
$X_0$ and $X_{\rm c}$ are identical.)
{\bf b} Relative changes in the $^{37}$Cl and $^{71}$Ga capture rates
(solid and dashed lines, respectively) and the flux of
$^8$B neutrinos (dot-dashed line)
}\label{properties}
\end{figure}

Changes in static solar models in response to opacity
modifications at various locations are illustrated in
Fig.~\ref{statdiff}, in terms of differences ($\deltar$) between the
modified and reference models at fixed fractional radius $r/R$.
Corresponding results for evolution models are shown in 
Fig.~\ref{evoldiff}.
The changes in the initial hydrogen abundance $X_0$
and mixing-length parameter $\alphac$ required to calibrate
the models are shown in Fig.~\ref{properties} as functions of $\log T_0$,
together with other overall properties of the models.
It is apparent that
$\deltar\ln T$, $\deltar \ln c$  and $\deltar \ln \rho$ 
behave almost like step functions at the location 
$T$ = $T_0$ of the imposed opacity change;
furthermore $\delta_r \ln T$ and $\delta_r \ln c$ are largely 
localized near this point.
On the other hand, $\deltar \ln p$ has a gentler behaviour.
These properties are investigated in more detail in Sect.~\ref{locmod} below.

Interestingly, the central temperature $T_{\rm c}$ depends in a non-monotonic
fashion on $\log T_0$, $\delta \ln T_{\rm c}$ being even slightly
negative for $\log T_0 \simeq 7.0$ (see Fig.~\ref{properties}a).
The reason for this behaviour is not obvious, although it
may be related to the steep decrease in $\deltar \ln T$ with
increasing $T$, and hence decreasing $r$, below the region of
immediate opacity modification (see also Sect.~\ref{locmod}).
The generation of the
high-energy neutrinos is strongly coupled to $T_{\rm c}$, 
and indeed the changes in the neutrino fluxes closely follow
$\delta \ln T_{\rm c}$ (cf. Fig.~\ref{properties}b).
In particular, there is also a 
slight decrease in the neutrino flux for
an opacity increase near $\log T_0 \simeq 7$.
This is in contrast to the calculation of Bahcall {\etal} (1969)
where the flux in models with increased opacity
is higher than in the reference model
at all values of $T_0$ (cf. Fig.~1 of their paper).
However, we note that Bahcall {\etal} 
used an opacity modification
with a width in $\log T$ approximately six times as large as in our case;
this is likely to have suppressed the finer structure in the
dependence of the neutrino flux on $T_0$.

Opacity increases generally lead to a decrease in the hydrogen abundance.
Indeed, an increase in opacity would tend to decrease the luminosity;
it follows from homology scaling (e.g. Kippenhahn \& Weigert 1990)
that this can be compensated for
by an increase in the mean molecular weight, i.e., a decrease in $X_0$.
When the opacity change is localized to the outer parts of the
radiative region, the changes in the core of the model, and
hence in the value of $X_0$ required for calibration, become very small.

\clearpage

Fig.~\ref{properties}, and a comparison of Figs~\ref{statdiff}
and \ref{evoldiff}, show substantial differences
between the static and the evolution models
in the changes in $X$ in the core of the model.
In the static case, $\deltar X$ is determined by the scaling
in Eq.~(\ref{Xscale}), together with the change in mass fraction $q$
at fixed $r$.
The overall effect is that the change is somewhat smaller in the
core than elsewhere.
For the evolution models, on the other hand, the change in
the initial abundance is modified by the change in the rate
of nuclear burning, which in turn depends predominantly on the change in
temperature.
As a result, the change in the central hydrogen abundance $X_{\rm c}$
generally reflects the change in the central temperature $T_{\rm c}$,
as is evident from Fig.~\ref{properties}.
This difference between static and evolution models 
in the behaviour of the hydrogen abundance in the core
leads to differences in the core response for other model
variables, as is clear from a comparison of Figs~\ref{statdiff}
and \ref{evoldiff}; on the other hand, there is striking similarity between
the results for static and evolution models elsewhere,
indicating that the use of static models is indeed a useful 
and computationally efficient technique for investigating the
response of the models to changes in the physics.

The behaviour of $\deltar\ln p$ and $\deltar\ln \rho$,
particularly in the convection zone, shows a striking variation with $T_0$,
the differences in the outer parts of the model changing sign
for $\log T_0 \simeq 6.74$.
As discussed in Sect. 4.2 below, these changes are related
to the complex behaviour of $\alphac$ and result from the
matching between the convection zone and the radiative interior,
and the calibration of the model to fixed radius.

We finally note from Fig.~\ref{properties} that,
as expected, the model differences are
negligible for opacity modifications confined essentially to
the convection zone, corresponding to $\log T_0 \lwig 6.25$;
this evidently follows from the fact that the structure of
the nearly adiabatic part of the convection zone is 
independent of opacity.

\section{Analysis of the model response}
In general, the response of the model to changes in aspects
of the physics is quite complex, as is also apparent from
the results presented so far.
However, it is possible to obtain some understanding of 
important aspects of the model changes from relatively simple analyses.
Here we consider the response in the vicinity of the imposed
opacity change, and the behaviour of changes in the convection zone.

\subsection{Local response of temperature and pressure}
\label{locmod}
The numerical results presented in Sect.~\ref{moddif} indicate a great
deal of regularity in the response near the point where the 
opacity modification is made.
Some properties of this behaviour can be understood quite simply
in terms of the perturbed equations of stellar structure.
For simplicity, we assume that the change in local mass and luminosity
can be neglected; the latter assumption is certainly satisfied
outside the nuclear-burning region, where the luminosity is fixed
by the calibration,
while we have found from the numerical results that the perturbation
in mass is comparatively small.
Also, we evidently consider only the radiative part of the star.
Under these assumptions the model response is
determined by just the equations 
of hydrostatic equilibrium and radiative energy transport which,
upon linearization, yield
\be
{\dd  \over \dd r} {\deltar p \over p} = 
{\dd \ln p \over \dd r} \left( {\deltar \rho \over \rho} -
{\deltar p \over p} \right) \; ,
\label{difhyd}
\ee
and
\bearr
{\dd  \over \dd r} {\deltar T \over T} = 
{\dd \ln T \over \dd r} 
\biggl[ \left( {\delta \kappa \over \kappa} \right)_{\rm int}
+ (\kappa_\rho + 1) {\deltar \rho \over \rho}
\nonumber \\
+ (\kappa_T - 4) {\deltar T \over T} \biggl]\; .
\label{difrad}
\eearr
In Eq.~(\ref{difrad})
$\kappa_\rho = (\partial \ln \kappa / \partial \ln \rho)_T$
and $\kappa_T = (\partial \ln \kappa / \partial \ln T)_\rho$,
and $( \delta \kappa / \kappa )_{\rm int}$ is the intrinsic 
opacity modification, corresponding to Eq. (\ref{kapmod}).
In Eq.~(\ref{difrad}) we neglected the effect 
on the opacity of the change in composition,
required by the calibration of the model;
this change is also found to be small, compared with the dominant effects.
Still neglecting effects of composition and assuming the ideal
gas law, the modifications in pressure, density and temperature
are related by
\be
{\deltar p \over p} =
{\deltar \rho \over \rho} + {\deltar T \over T}  \; .
\label{difeos}
\ee

It is evidently possible to carry out a complete description
of the modifications to the model through numerical solution 
of these equations,
with appropriate boundary conditions;
the analysis should then include also the changes in mass and luminosity,
as described by the linearized versions of the relevant equations,
and the effect on the composition.
Here, however, we approximate the equations further, in order
to obtain a rough analytical solution which may be used
to interpret the numerical results.
{}From Eq.~(\ref{difrad}) it is evident that an intrinsic modification
$( \delta \kappa / \kappa )_{\rm int}$ approximating
a delta function induces a step function in $\deltar T/T$,
while, from Eqs (\ref{difhyd}) and (\ref{difeos}),
$\deltar p/p$ results from an integration over this step function.
Thus it is plausible, as confirmed by the numerical results,
that near the modification $\deltar T/T$ is substantially
larger than $\deltar p/p$.
Consequently we neglect $\deltar p/p$ in Eq. (\ref{difeos}) to
obtain $\deltar \rho/\rho \simeq - \deltar T/T$ and hence,
from Eq.~(\ref{difrad}),
\be
{\dd  \over \dd \ln T} {\deltar T \over T} 
+ \zeta {\deltar T \over T} =
\left( {\delta \kappa \over \kappa} \right)_{\rm int} \; ,
\label{difdelt}
\ee
where $\zeta = 5 + \kappa_\rho - \kappa_T$.
We assume that $\kappa_\rho$ and $\kappa_T$ are approximately constant,
and take as boundary condition that $\deltar T /T = (\deltar T/T)_{\rm s}$
at a suitable reference temperature $T_{\rm s} < T_0$.
Then the solution to Eq.~(\ref{difdelt}) is
%
\be
{\deltar T \over T} ~=~ \left\{
\begin{array}{ll}
\dis\left({T_{\rm s} \over T} \right)^\zeta
\left( {\deltar T \over T} \right)_{\rm s} & 
\hskip -1cm {\rm for} \quad T < T_0 \; , \\
\dis\left({T_{\rm s} \over T} \right)^\zeta
\left( {\deltar T \over T} \right)_{\rm s} +
K_0 \left({T_{\rm 0} \over T} \right)^\zeta & \\
 & \hskip -1cm {\rm for} \quad T > T_0 \; , \\
\end{array}
\right.
\label{soldelt}
\ee
where
\be
K_0 
= \int \left( {\delta \kappa \over \kappa} \right)_{\rm int} \dd \ln T  \; .
\label{condelt}
\ee
The exponent $\zeta$ is large:
at a typical point in the solar radiative interior,
$\kappa_T \simeq -2.5$ and 
$\kappa_\rho \simeq 0.5$, so that $\zeta \simeq 8$.
Consequently, the term in $(T_{\rm s}/T)^\zeta$ decreases
very rapidly with increasing depth and, as a first
approximation, $\deltar T/T \simeq K_0 (T_0/T)^\zeta$
for $T > T_0$ and zero otherwise.

To estimate $\deltar p/p$ we write Eq.~(\ref{difhyd})
as, using Eq.~(\ref{difeos}),
\be
{\dd  \over \dd \ln T} {\deltar p \over p} = 
- \nabla^{-1} {\deltar T \over T} \; ,
\label{difdelp}
\ee
where $\nabla = \dd \ln T / \dd \ln p$.
Assuming $\nabla$ to be roughly constant over the relevant region
and using the approximate solution for $\deltar T/T$, we obtain
\be
{\deltar p \over p} \simeq \left\{
\begin{array}{ll}
\dis\left({\deltar p \over p}\right)_0 &
{\rm for} \quad T < T_0 \; , \\
\dis\left({\deltar p \over p}\right)_0 + {K_0 \over \zeta \nabla} 
\left[ \left({T_0 \over T} \right)^\zeta - 1 \right] &
{\rm for} \quad T > T_0 \; , \\
\end{array}
\right.
\label{soldelp}
\ee
where
$(\deltar p / p)_0$ is the value of $\deltar p / p$ at $T = T_0$.

\begin{figure}
\vspace{0cm}
\hspace{0cm}\epsfxsize=8.8cm \epsfbox{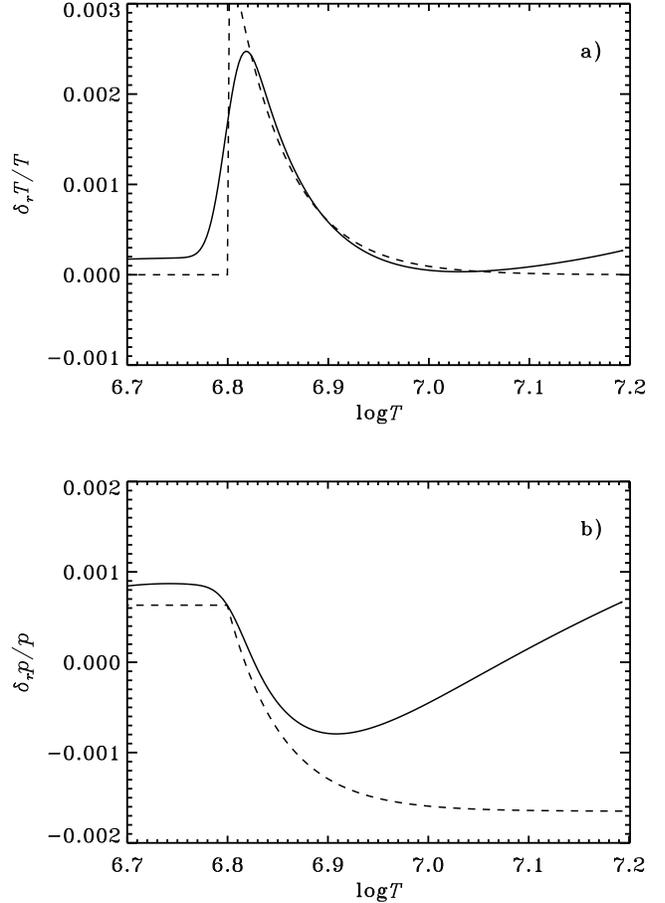}
\vspace{0cm}
\caption[]
{Comparisons of actual model differences (solid lines)
with analytical approximations (dashed lines),
for a static model with $\log T_0 = 6.8$,
$A = 0.02$ and $\Delta = 0.02$;
to illustrate the behaviour of the solution,
position in the model is indicated by $\log T$.
{\bf a} Relative difference $\deltar T/T$ in temperature;
the analytical solution was obtained from Eqs~(\ref{soldelt}) and
(\ref{condelt}), neglecting the term in $(\delta_r T/T)_{\rm s}$.
{\bf b} Relative difference $\deltar p/p$ in pressure;
the analytical solution was obtained from Eqs~(\ref{soldelp}),
fitting $(\deltar p/p)_0$ to the numerical solution
at the location $T = T_0$ of the opacity modification
}\label{testanal}
\end{figure}

To test these expressions, we consider in Fig.~\ref{testanal}
the case, already presented in Fig.~\ref{statdiff}b,
of a static model modified with $\log T_0 = 6.8$;
since the analytical solutions depend predominantly
on temperature, we show the results as a function of $\log T$,
rather than $r/R$.
The analytical solution for $\deltar T/T$ is clearly in good agreement
with the numerical results; in particular, the rapid decrease,
as $T^{-\zeta}$, for $T > T_0$ accounts for the localized
nature of the temperature modification.
The fit is rather less satisfactory for $\deltar p/p$,
although the analytical solution approximately recovers
the magnitude of the variation in the vicinity of $T = T_0$;
in fact, it is plausible that the neglect of the finite
extent of $(\delta \kappa/\kappa)_{\rm int}$,
the neglect of the change in composition and the assumption that
$\nabla$ is constant are of doubtful validity in this case.
Even so, the analytical solutions do provide some insight into
the nature of the model response.

We finally note that the change in sound speed is given by
\setlength{\arraycolsep}{1mm}
\bearr
\deltar\ln c^2
 & = &\deltar \ln \gamma_1 + \deltar \ln \left( {p \over \rho}\right) 
\nonumber \\
& \simeq & \deltar \ln \gamma_1 + \deltar \ln T - \deltar \ln \mu \; ,
\label{lnc}
\eearr
where in the last equality the ideal gas law was assumed;
here $\gamma_1 \equiv (\partial \ln p / \partial \ln \rho)_s$,
the derivative being taken at constant specific entropy $s$, 
and $\mu$ is the mean molecular weight.
Outside ionization zones of abundant elements, $\deltar \ln \gamma_1$
can be approximately neglected; also, $\deltar \ln \mu$ is almost constant,
except in the core where the composition has been altered by nuclear burning.
As a result, as a first approximation $\deltar \ln c^2$ 
is obtained from $\deltar \ln T$ by a constant shift,
as is indeed observed in Figs~\ref{statdiff} and \ref{evoldiff}.

\subsection{Model differences in the convection zone}
\label{conmod}
The bulk of the convection zone is very nearly adiabatically stratified;
hence its structure depends only on the equation of 
state, the composition and the (constant) specific entropy $s$.
Indeed, it is evident from Fig.~\ref{properties} that the
model is insensitive to the opacity in the convection zone.
Assuming that the physics of the atmosphere is given,
$s$ is determined, within mixing-length theory,
by the mixing-length parameter $\alphac$.
If, as in the calculations presented in Sect.~3,
we also assume 
that the equation of state and the heavy-element abundance $Z$ are fixed,
the structure of the adiabatic part of the convection zone is then
essentially characterized by $\alphac$ 
and the envelope hydrogen abundance $\Xs$.
Some simple properties of such convective envelopes are
discussed in the Appendix.

\begin{table*}
\caption[]{Properties of the envelope models.
$\Xs$ is the envelope hydrogen abundance, $\pcz$
is the pressure at the base of the convection zone
and $K$ is defined by the adiabatic relation (\ref{prho})} 
\begin{flushleft}
\begin{tabular}{cccccc}
\hline
\hline
Model No.&$\Xs$&$\alphac$&$\dcz/R$&$\pcz$&$K$\\
         &     &         &        &$(\dyn \cm^{-2})$&(cgs)\\
\hline
En1&0.697867&1.81586&0.274105&$4.44746\;10^{13}$&$9.52969\;10^{14}$\\
En2&0.697867&1.84273&0.277000&$4.75692\;10^{13}$&$9.33427\;10^{14}$\\
En3&0.702867&1.81586&0.273757&$4.34019\;10^{13}$&$9.65974\;10^{14}$\\
\hline
\end{tabular}
\end{flushleft}
\end{table*}

\begin{figure}
\vspace{0cm}
\hspace{0cm}\epsfxsize=8.8cm \epsfbox{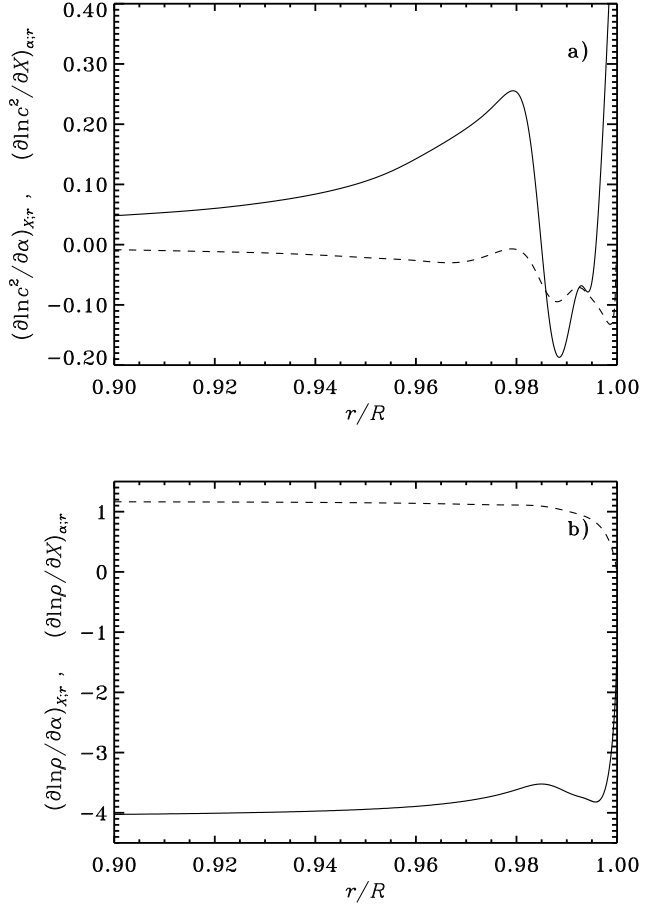}
\vspace{0cm}
\caption[]
{Estimated partial derivatives of model quantities
in the outer parts of the convection zone, with respect
to the envelope hydrogen abundance $\Xs$ (solid lines)
and the mixing-length parameter $\alphac$ (dashed lines).
{\bf a} Derivatives of squared sound speed.
{\bf b} Derivatives of density}
\label{parder}
\end{figure}

For model changes so small that the
linear approximation is valid, the difference in any quantity $\phi$
may be expressed in terms of the changes $\delta \Xs$
and $\delta \alphac$ in $\Xs$ and $\alphac$, as
\be
\deltar \phi 
\simeq \left({\partial \phi\over \partial \Xs}\right)_{\alphac;r} \delta \Xs
+ \left({\partial \phi\over \partial \alphac}\right)_{\Xs; r} \delta \alphac
\; .
\label{env}
\ee
The partial derivatives in Eq.~(\ref{env}) may be estimated
from differences between envelope models differing only 
in $\Xs$ or $\alphac$ as, for example
$(\partial \phi / \partial \Xs)_{\alphac;r} \simeq \deltar \phi / \delta \Xs$.
Here we consider three envelope models 
constructed using the same input physics 
as for the complete models described earlier,
but with prescribed values of $\Xs$ and $\alphac$.
Some properties of the models are listed in Table~2.
In particular, $K$ is the value of the constant in the
adiabatic relation between $p$ and $\rho$
[cf. Eq. (\ref{prho}) in the Appendix],
averaged over the lower part of the convection zone, with $r \le 0.8 R$.
The resulting derivatives, in the outer parts of the
convection zone, are shown in Fig.~\ref{parder};
at greater depth the derivatives of $\ln c^2$ are very small, 
and the derivatives of $\ln \rho$ are essentially constant,
in accordance with the discussion in the Appendix
[cf. in particular Eq.~(\ref{deltap})].

\begin{figure}
\vspace{0cm}
\hspace{0cm}\epsfxsize=8.8cm \epsfbox{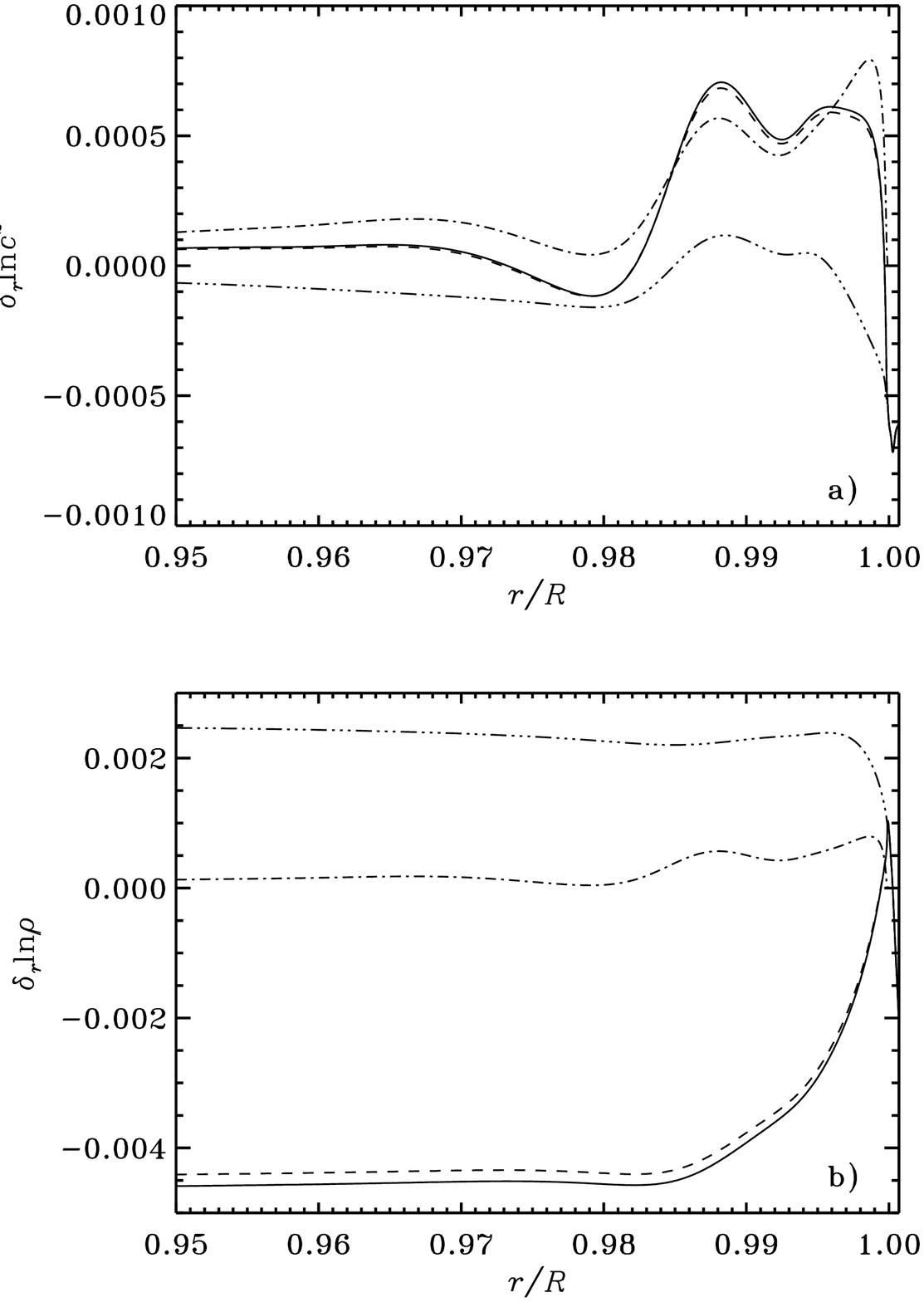}
\vspace{0cm}
\caption[]
{Differences in squared sound speed (panel a)
and density (panel b) in the upper part of
the convection zone, for a static model modified at $\log T_0 = 7.0$.
The solid line shows the computed difference, from Fig.~\ref{statdiff}a,
the dash-dot and dash-triple-dot lines show the
contributions in Eq.~(\ref{env}) from $\delta \alphac$ and $\delta \Xs$,
respectively, and the dashed line shows the sum of these contributions}
\label{ccon}
\end{figure}

\begin{figure}
\vspace{0cm}
\hspace{0cm}\epsfxsize=8.8cm \epsfbox{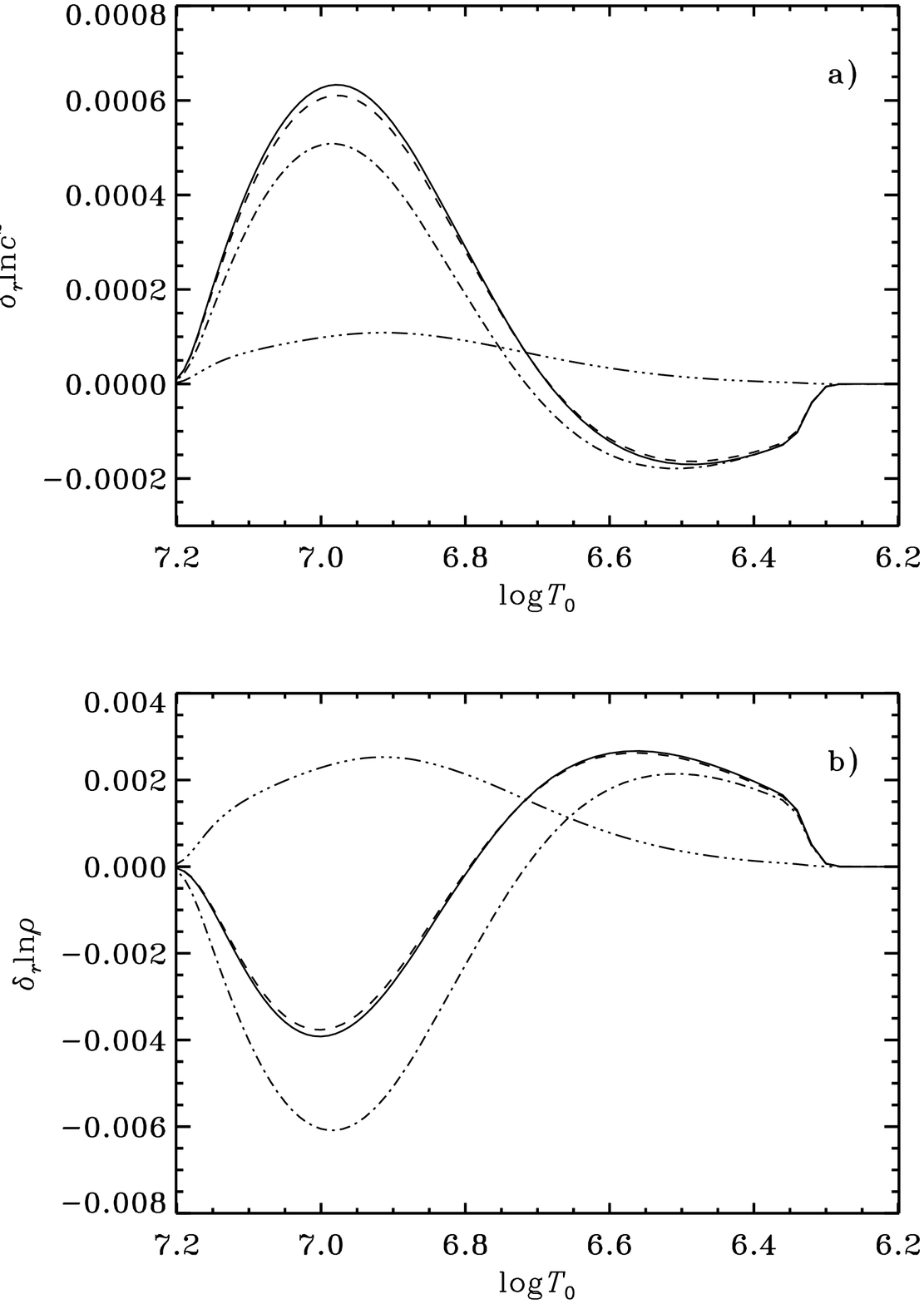}
\vspace{0cm}
\caption[]
{Differences in squared sound speed (panel a) and
density (panel b) at $r = 0.99 R$ in modified static models,
as a function of the location $\log T_0$ of the opacity modification.
The same line styles as in Fig.~\ref{ccon} have been used}
\label{comcon}
\end{figure}

For the models computed with changes in the internal
opacity, the changes in $\alphac$ and $\Xs$ arise
from the calibration to fixed radius and luminosity.
More precisely, $\Xs$ is essentially determined by 
the condition that the luminosity is fixed and, given $\Xs$, 
$\alphac$ is determined such as to ensure a continuous match
at the base of the convection zone, with a fixed surface radius.
Given the partial derivatives determined from the
envelope calculations, we may test Eq.~(\ref{env}) 
by comparing actual differences found in Sect.~3,
resulting from opacity modifications,
with those obtained from Eq.~(\ref{env})
with the changes in $\Xs$ and $\alphac$ shown in Fig.~\ref{properties}a.
Results for static models
are shown in Figs~\ref{ccon} and \ref{comcon}.
It is evident from Fig.~\ref{ccon}, for $\log T_0 = 7.0$,
that the simple relation (\ref{env})
reproduces quite accurately the details of the sound-speed changes in the 
hydrogen and helium ionization zones.
Also, the dominant contribution in this case clearly results from
the change in $\alphac$.

Fig.~\ref{comcon} shows the changes in squared sound speed 
and density at $r = 0.99 R$, as well as the contributions
from the changes in $\alphac$ and $\Xs$, as a function of $\log T_0$.
The dependence of the individual contributions clearly follows
the variation of $\delta \alphac$ and $\delta \Xs$, 
shown in Fig.~\ref{properties}a.
In particular, the complex behaviour of $\delta \alphac$ is reflected in
a corresponding behaviour of $\deltar \ln c^2$ and
$\deltar \ln \rho$, including the changes in sign
for $6.7 \le \log T_0 \le 6.8$.
For opacity changes closer to
the base of the convection zone, the model differences are
dominated by the change in $\alphac$, the contributions from
the change in $\Xs$ becoming negligible.
The main reasons for these variations are not obvious;
however, it seems likely that the behaviour 
in the radiative interior
(cf. Fig.~\ref{statdiff}) causes the change of sign in
of $\delta_r p/p$ and $\delta_r \rho/\rho$ 
at the base of the convection zone, as the location of
the opacity modification moves closer to the surface.
Since $\delta_r p/p$ is directly linked to the change in the
polytropic constant $K$ by Eq. (\ref{deltap}),
and hence to the change in $\alphac$ by Eq. (\ref{numder})
this is the probable cause of the change in $\alphac$ and hence,
by Eq. (\ref{deldcznum}), in the depth of
the convection zone (see also Fig.~\ref{properties}).
In turn, the depth of the convection zone largely controls the behaviour
of $\deltar \ln T$ and $\deltar \ln c^2$
in the outer parts of the radiative region:
For $\log T_0 \gwig 6.74$, the convection zone is shallower
in the modified model than in the reference model.
Consequently, the temperature and sound-speed gradients are less steep
in the modified model just beneath the convection zone;
since, as argued in the Appendix, the sound-speed difference is
very small in the lower parts of the convection zone,
this leads to a negative sound-speed difference in that part of
the radiative interior which lies outside the opacity modification.
This trend is reversed for $\log T_0 \lwig 6.74$.

\section{Tests of the opacity kernels}
As indicated by Eq. (1),
the sensitivity of a physical quantity to the opacity perturbations
can be measured by opacity kernels.
In this section we present examples of such kernels.
We furthermore test the accuracy with which kernels
can reproduce the results of a large opacity change,
by applying them to
solar models with artificially reduced opacity values 
which simulate the effect of Weakly Massive Interacting Particles (WIMPs)
(Christensen-Dalsgaard 1992).

\begin{figure}
\vspace{0cm}
\hspace{0cm}\epsfxsize=8.8cm \epsfbox{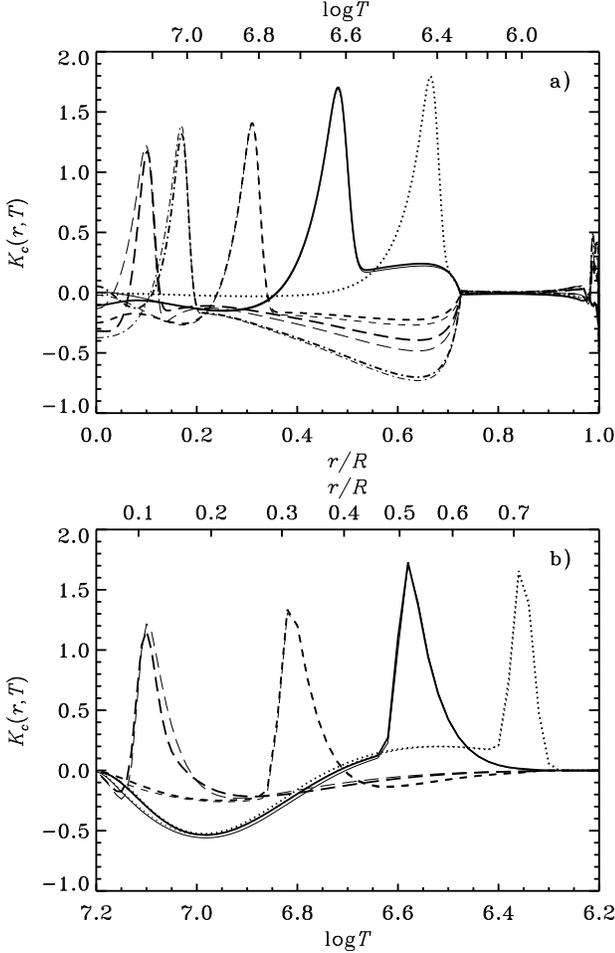}
\vspace{0cm}
\caption[]{
Kernels $K_c(r, T)$ relating the change in opacity to the
sound-speed change (cf. Eq. \ref{kernelc}).
Heavy lines show kernels based on evolution calculation,
while the thin lines were obtained from static models.
{\bf a} Kernels as a function of fractional radius $r/R$
for $\log T = 6.4$ (dotted line), 
$\log T = 6.6$ (solid line),
$\log T = 6.8$ (short-dashed line),
$\log T = 7.0$ (dot-dashed line), and
$\log T = 7.1$ (long-dashed line).
The upper abscissa shows $\log T$, corresponding to the lower abscissa.
{\bf b} Kernels as a function of $\log T$, 
for $r/R = 0.70$ (dotted line),
$r/R = 0.50$ (solid line),
$r/R = 0.30$ (short-dashed line), and
$r/R = 0.10$ (long-dashed line).
The upper abscissa shows $r/R$, corresponding to the lower abscissa
}\label{cker}
\end{figure}

Fig. 9 shows opacity kernels $K_c(r, T)$ for sound speed,
defined such that
\be
\left({\delta_r c \over c}\right) (r)
= \int K_c(r, T) \delta \log \kappa(T) \dd \log T \; .
\label{kernelc}
\ee 
These kernels were determined from the model changes discussed
in Sect.~3.2 and are illustrated as functions of $r$ and of $T$.
Results for both static and evolution models are shown.
As discussed in Sect.~4, the kernel for a given temperature
$T$ is localized in $r$ fairly
close to the position in the model corresponding to that temperature.
Also, it is interesting that the kernels can be determined
quite accurately from static models, except in the core
where the evolution of the hydrogen abundance is substantially
affected by the opacity modification.
This provides some justification for the procedure adopted by
Elliott (1995), which neglected possible evolutionary effects.

To illustrate the effects on important global properties of the model
Fig.~10 shows kernels for the neutrino fluxes $\Phi_\nu$
as well as the small frequency separation $6 D_0$ (cf. Eq. 6).
Due to the high sensitivity of the neutrino fluxes to
the opacity modifications, the neutrino kernels are defined
specifically by
\be
\delta \ln \Phi_\nu = \int K_\Phi(T) \delta \log \kappa(T) \dd \log T \; .
\ee

\begin{figure}
\vspace{0cm}
\hspace{0cm}\epsfxsize=8.8cm \epsfbox{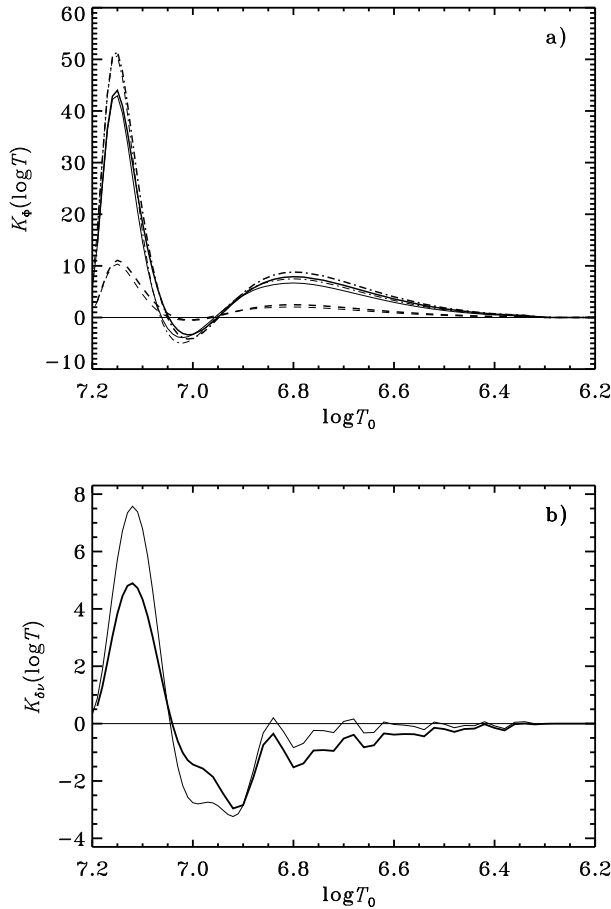}
\vspace{0cm}
\caption[]{
Kernels for global model quantities.
Heavy lines show kernels based on evolution calculation,
while the thin lines were obtained from static models.
{\bf a} Kernels for logarithmic changes in the $^{37}$Cl and $^{71}$Ga 
capture rates (solid and dashed lines, respectively) and in the flux of
$^8$B neutrinos (dot-dashed line) (cf. Eq. 20).
{\bf b} Kernels for the averaged small separation $6 D_0$ 
(cf. Eqs \ref{freqsep} and \ref{d0})}
\label{globker}
\end{figure}

\begin{table}
\caption{Properties of solar models with reduced core opacity,
as well as the corresponding reference evolution model EW0}
\begin{flushleft}
\begin{tabular}{ccccccc}
\hline
\hline
&&\multicolumn{2}{c}{$^{37}$Cl neutrino flux}
      &\multicolumn{2}{c}{6$D_0$}&\\  
&&\multicolumn{2}{c}{(SNU)}
      &\multicolumn{2}{c}{($\mu$Hz)}&\\  
Model&$A$&Computed&kernels&Computed&kernels&\\
\hline
EW0&0.0&8.89& -- &9.169& --  &\\
 & & & & & &\\
SW1&0.2&4.87&4.76&8.023&8.071&\\
SW2&0.4&2.96&2.55 &7.034&7.105&\\
EW1&0.2&4.59&4.61&8.395&8.439&\\
EW2&0.4&2.67&2.39&7.697&7.767&\\
\hline
\end{tabular}
\end{flushleft}
\end{table}

It is evident that these kernels correspond precisely, apart from
a scaling, to the neutrino-flux differences shown in Fig. 4b.
Also, it is striking that even for the neutrino flux there is
little difference between the results for static and evolution models.
In contrast, kernels for the small frequency separation, which depends strongly
on the composition gradient in the core, cannot be estimated 
accurately from the static models.

\begin{figure}
\vspace{0cm}
\hspace{0cm}\epsfxsize=8.8cm \epsfbox{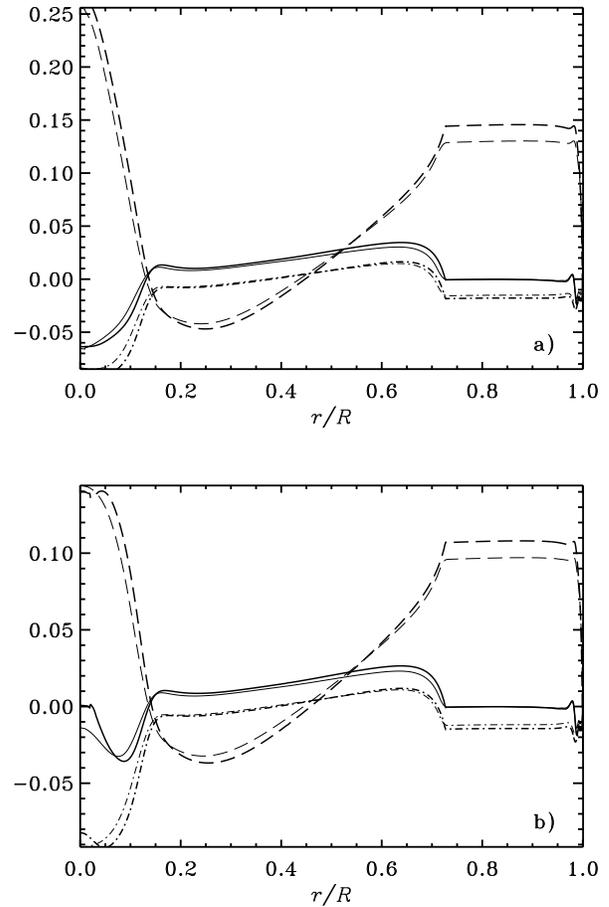}
\vspace{0cm}
\caption[]
{Differences at fixed fractional radius $r/R$ 
between models with opacity modification to simulate the
effect of WIMPs (cf. Christensen-Dalsgaard 1992)
and the corresponding reference models,
in the sense (modified) -- (reference).
The thick lines represent the values obtained from the kernels while 
the thin lines show the original differences.
Variables shown are $\deltar \ln c$ (solid line); $\deltar \ln \rho$ 
(dashed line); $\deltar \ln T$ (dash-dot line).
Some properties of the models are provided in Table 3.
{\bf a} Differences between the modified 
static model SW2 and the reference model EW0.
{\bf b} Differences between the modified evolution 
model EW2 and the reference model EW0}
\label{figwimp}
\end{figure}

To test these kernels, we use the models 
of Christensen-Dalsgaard (1992), with reduced core opacity.
Specifically, the opacity
modification was determined from
\be
{\log}~\kappa = {\log}~\kappa_0~-~A f(\log T) \; ,
\ee
where 
\[
f(\log T)~=~\left\{ \begin{array}{ll}
 \exp \left[ -\dis\left({\log T - \log T_1 
 \over \Delta \log T}\right)^2 \right] 
\dis, 
   & \mbox{if $T < T_1$}\\
1 & \mbox{otherwise;}
\end{array}\right.
\]
\be
\vspace{-0.5cm}
\qquad
\ee
as did Christensen-Dalsgaard (1992),
we used $\Delta \log T = 0.04$ and $\log T_1 = 7.1$.
This form of $f$ provides an opacity decrease over a well defined region in 
$\log T$ with a continuous transition to zero at lower temperature. 
(Note that in these models the unmodified opacity $\kappa_0$ was obtained using 
the Los Alamos Opacity Library of Huebner {\etal} 1977.)
It was found that
the neutrino flux was reduced by a factor of more than three for $A = 0.4$
(see also Table~3). 

Table 3 compares the values of the ${}^{37}$Cl neutrino
flux and the small frequency separation computed for the models with
reduced opacity with those estimated by means of the
kernels shown in Fig.~10.
It is evident that there is good agreement between the
values of $D_0$ for both $A = 0.2$ and 0.4. 
The agreement in neutrino flux is somewhat better for $A = 0.2$, the 
error being $10 - 15$ \% for $A = 0.4$. 
However,
it is remarkable that linearized expressions of the form given in
Eq. (\ref{kernel}) remain reasonably precise for a reduction in
opacity by more than a factor of two.

To test the precision with which the kernels reproduce the
structure of the model, Fig.~11 compares differences in $c^2$ and
$\rho$ obtained from the kernels in Fig. 9 and the corresponding
density kernels, with the actual model differences.
Results are shown both for the static model SW2 and the
evolution model EW2, both corresponding to $A = 0.4$.
As before, the agreement is quite close  
both for evolution and for static models.  

\section{Conclusion}
 Our understanding of the solar internal structure has improved
significantly over the last few years due to the rapid progress
in helioseismology as well as in the construction of solar
models with improved input physics.
Although much of the knowledge has been obtained
through inverse analysis, the starting point is always 
solar models; thus it is important to investigate the
sensitivity of the solar structure to the input physics and to
determine the presence of any region which is particularly
sensitive to a specific parameter. Since the major uncertainties
in the microphysics come from the opacity, we have focussed on the
examination of the solar structure by means of a localised
opacity change as a function of temperature.


The sensitivity of the solar structure was represented by
kernels relating the opacity changes to neutrino flux, frequency separation
at low degree as parametrised by $D_0$ and the structure difference between 
modified and reference models. These kernels were subsequently used to 
derive the same parameters corresponding to a reduction in opacity by a 
factor of more than two in the models of Christensen-Dalsgaard (1992),
simulating the effects of Weakly Interacting Massive Particles.
We found that the kernels were remarkably successful in estimating
the changes in the solar structure caused by 
even such a large change in the input physics.

A natural next step in this investigation is
to study systematically how the oscillation 
frequencies are affected by opacity changes and also how well 
these frequency changes can be reproduced by kernels.
Furthermore, in a preliminary analysis
Tripathy, Basu \& Christensen-Dalsgaard (1997)
found that much of the current discrepancy between the
helioseismically inferred solar sound speed and the
sound speed of a standard solar model can be understood
in terms of modest modifications to the opacity.
More detailed analyses of this nature are now under way.

\begin{acknowledgements}
We thank S. Basu and J. Elliott for useful discussion
as well as for comments on the manuscript.
JC-D is grateful to the High Altitude Observatory for
hospitality during much of the preparation of the paper.
The work was supported in by Danish Natural Science Research Council,
and by the Danish National 
Research Foundation through its establishment of the 
Theoretical Astrophysics Center.
SCT acknowledges financial support from the
Department of Science and Technology, Government of India.
\end{acknowledgements}

\appendix
\section{Some properties of convective envelopes}

For convenience, we present some simple approximate properties
of convective envelopes, which are useful for the interpretation
of the numerical results in Sect.~4.2.
Further details were provided by Christensen-Dalsgaard (1997).
Note also that Baturin \& Ayukov (1995, 1996) carried out a careful
analysis of the properties of convective envelopes
and their match to the radiative interior.

In the convection zone the stratification departs substantially
from being adiabatic only in a very thin region just below the photosphere.
Thus the structure of the bulk of the convection zone is determined 
by the equation of hydrostatic equilibrium 
\be
{\dd p\over \dd r} = - {G m \rho\over r^2} 
\label{hydro}
\ee 
($G$ being the gravitational constant and $m$ the mass interior to $r$),
and the relation 
\be
{1\over \gamma} \equiv {\dd \ln \rho \over \dd \ln p} 
\simeq {1\over \gamma_1} \; .
\label{gamma}
\ee
To this approximation, the convection-zone structure 
therefore depends only on the equation of state, the composition and 
the constant specific entropy $s$, the latter being fixed by 
the mixing-length parameter $\alphac$.
In practice, a more useful characterization of the bulk of the
convection zone follows from noting that
$\gamma_1$ is approximately constant
outside the dominant ionization zones of hydrogen and helium;
then Eq.~(\ref{gamma}) shows that $p$ and $\rho$ are related by
\be
p \simeq K \rho^{\gamma_1} \; ,
\label{prho}
\ee
where the constant $K$ is closely related to $s$.
Assuming $\gamma_1$ again to be constant and neglecting the mass in
the convection zone, it is readily shown from
Eqs~(\ref{hydro}) and (\ref{gamma}) that
$u \equiv p/\rho = c^2/\gamma_1$ is given by
\be
u \simeq G M  \left( 1 - {1 \over \gamma_1} \right)
\left({1 \over r} - {1 \over R^*} \right) \; ,
\label{solu}
\ee
approximately valid in the lower parts of the convection zone
(e.g. Christensen-Dalsgaard 1986;
Dziembowski, Pamyatnykh \& Sienkiewicz 1992;
Christensen-Dalsgaard \& D\"appen 1992; Baturin \& Mironova 1995);
here $M$ is the mass of the model and $R^* \simeq R$, its surface radius.
{}From Eqs~(\ref{prho}) and (\ref{solu}) we also obtain
\be
p^{1 - 1/\gamma_1} \simeq
K^{-1/\gamma_1} {\gamma_1 -1 \over \gamma_1} 
G M \left( {1 \over r} - {1 \over R_*} \right) \; .
\label{solp}
\ee

Eq.~(\ref{solu}) indicates that $u$, and therefore $c$,
depend little on the details of the physics of the model;
in particular, if $M$ and $R$ are assumed to be fixed,
as in the case of calibrated solar models, $\delta_r u \simeq 0$.
This is confirmed by Figs~\ref{statdiff} and \ref{evoldiff},
which show that $\deltar \ln c$ is very small in the bulk of 
the convection zone.
{}From Eq.~(\ref{solp}) it furthermore follows that
\be
\deltar \ln p \simeq \deltar \ln \rho 
\simeq - {1 \over \gamma_1 - 1} \delta \ln K 
\label{deltap}
\ee
are approximately constant, again in accordance
with Figs~\ref{statdiff} and \ref{evoldiff}.
Finally, assuming the ideal gas law, we have that
\be
\deltar \ln T \simeq  \deltar \ln \mu \; .
\label{deltat}
\ee

These relations may be used to investigate the changes in
the model resulting from changes in the convection-zone parameters.
Of particular interest are conditions at the base of the convective
envelope, defining the match to the radiative interior.
Neglecting possible convective overshoot,
$\nablaad = \nablarad$, the radiative temperature gradient, at this point.
For calibrated models the luminosity $L$ is unchanged;
then $\nablarad \propto \kappa p / T^4$.
Assuming also that $\nablaad$ is constant
and that the heavy-element abundance $Z$ is fixed,
we find that the change in the pressure $\pcz$ at the base
of the convection zone is related to the changes in $K$
and the envelope hydrogen abundance $\Xs$ by
\bearr
\delta \ln \pcz & \simeq & - {4 - \kappat_T \over (4 - \kappat_T) (\gamma_1 - 1)
- \gamma_1(\kappat_p + 1)} \, \delta \ln K \nonumber \\
&  & - {\gamma_1 [ (4 - \kappat_T) \mu_X - \kappat_X] 
\over (4 - \kappat_T) (\gamma_1 - 1)
- \gamma_1(\kappat_p + 1)} \, \delta \ln \Xs 
\label{delpcz}
\eearr
(see also Christensen-Dalsgaard 1997);
here 
\bearr
\kappat_p = (\partial \ln \kappa / \partial \ln p)_{T,X}\; , \qquad
\kappat_T = (\partial \ln \kappa / \partial \ln T)_{p,X}\;, \nonumber \\
\kappat_X = (\partial \ln \kappa / \partial \ln X)_{p,T}\;, \qquad
\mu_X = (\partial \ln \mu / \partial \ln X)_Z\; . \nonumber
\eearr
Having obtained $\delta \ln \pcz$, the change in the depth of
the convection zone can be determined from Eq.~(\ref{solp}) as
\be
\delta \ln \dcz \simeq {1 \over \gamma_1} {\rcz \over R}
[\delta \ln K + (\gamma_1 - 1) \delta \ln \pcz] \; ,
\label{deldcz}
\ee
where $\rcz$ is the radius at the base of the convection zone.
[Note that this relation differs from Eq.~(11) of Christensen-Dalsgaard (1997).
There it was assumed that the interior properties of the model
were largely unchanged while the surface radius was allowed to change;
this is the case relevant, for example, to the calibration of
solar models to have a specific radius.
Here we have kept $R$ fixed and the change in $\dcz$ corresponds to
changes in the properties of the radiative interior of the model.]

It is of some interest to compare these simple relations with the
numerical results obtained for the envelope models listed in Table 2.
To do so, we first need to relate the change in $K$ to the changes
in $\alphac$ and $\Xs$.
{}From the results in Table 2 
\be
\left({\partial \ln K \over \partial \ln \alphac } \right)_{\Xs} 
\simeq -1.40 \; , \qquad
\left({\partial \ln K \over \partial \ln \Xs } \right)_{\alphac} 
\simeq 1.90 \; .
\label{numder}
\ee
As discussed by Christensen-Dalsgaard (1997), the relation between
$\alphac$ and $K$, at fixed composition,
follows simply from the properties of the mixing-length theory;
the result of such an analysis, using the properties of the reference
envelope model En1, agrees quite closely with the value given above.
It is less straightforward to derive a simple expression
for the relation between $\delta \ln K$ and $\delta \ln \Xs$.

To evaluate the changes in $\pcz$ and $\dcz$, we need
the derivatives of $\kappa$ and $\mu$.
At the base of the convection zone in the reference envelope En1
we have the following values:
$\kappat_p = 0.63$, $\kappat_T = -3.70$, $\kappat_X = -0.16$, and
$\mu_X = -0.54$.
Thus, from Eq.~(\ref{delpcz}) we obtain
$\delta \ln \pcz \simeq -3.20 \delta \ln K + 2.99 \delta \ln \Xs$,
and hence, from Eq.~(\ref{deldcz}),
\be
\delta \ln \dcz \simeq -0.49 \delta \ln K + 0.87 \delta \ln \Xs \; .
\label{deldczknum}
\ee
Using also Eqs~(\ref{numder}) we find
\be
\delta \ln \dcz \simeq 0.69 \delta \ln \alphac - 0.07 \delta \ln \Xs \; .
\label{deldcznum}
\ee
For comparison, the results in Table 2 give
\be
\left( { \partial \ln \dcz \over \partial \ln \alphac}\right)_{\Xs} 
\simeq 0.72 \; , \qquad
\left( { \partial \ln \dcz \over \partial \ln \Xs}\right)_{\alphac} 
\simeq -0.18 \; .
\ee
Evidently, the $\alphac$-derivative is in reasonable agreement
with Eq.~(\ref{deldcznum});
although the agreement appears less satisfactory for the derivative
with respect to $\Xs$, it should be noticed that small coefficient
in Eq.~(\ref{deldcznum}) arises from near cancellation between
the contributions from the two terms in Eq.~(\ref{deldczknum}).

\bibliographystyle{plain}


\end{document}